\documentclass[twocolumn,aps,pra,superscriptaddress,longbibliography]{revtex4-1}
\usepackage{color}
\usepackage{amsmath}
\usepackage{amssymb}
\usepackage{graphicx}
\usepackage{tikz}
\usepackage{multirow}
\usepackage{float}
\usepackage{bbm}
\usepackage{dsfont}
\usepackage{cases}
\usepackage{mathtools}

\usepackage{epsfig}
\usepackage{verbatim}
\usepackage{array}
\usepackage{setspace}

\usepackage{scalerel}
\usepackage{stackengine,wasysym}
\usepackage{empheq, nccmath}

\usepackage[T1]{fontenc}

\newcommand\blue[1]{{\color{blue}#1}}

\usepackage[unicode=true,
bookmarks=true,bookmarksnumbered=false,bookmarksopen=false,
breaklinks=false,pdfborder={0 0 1},backref=false,colorlinks=true]
{hyperref}
\hypersetup{
	linkcolor=magenta, urlcolor=blue, citecolor=blue, pdfstartview={FitH}, hyperfootnotes=false, unicode=true}

\makeatletter
\@ifundefined{textcolor}{}
{%
	\definecolor{BLACK}{gray}{0}
	\definecolor{WHITE}{gray}{1}
	\definecolor{RED}{rgb}{1,0,0}
	\definecolor{GREEN}{rgb}{0,1,0}
	\definecolor{BLUE}{rgb}{0,0,1}
	\definecolor{CYAN}{cmyk}{1,0,0,0}
	\definecolor{MAGENTA}{cmyk}{0,1,0,0}
	\definecolor{YELLOW}{cmyk}{0,0,1,0}
}

\begin{document}

\preprint{APS/123-QED}

\title{Low-temperature series expansion of square lattice Ising model: A study based on Fisher zeros}

\author{De-Zhang Li}
\affiliation{Quantum Science Center of Guangdong-Hong Kong-Macao Greater Bay Area, Shenzhen, Guangdong 518045, China}
\author{Xin Wang}
\thanks{Corresponding author: x.wang@cityu.edu.hk}
\address {Department of Physics, City University of Hong Kong, Hong Kong SAR, China}
\affiliation{City University of Hong Kong Shenzhen Research Institute, Shenzhen, Guangdong 518057, China}
\affiliation{Quantum Science Center of Guangdong-Hong Kong-Macao Greater Bay Area, Shenzhen, Guangdong 518045, China}
\author{Xiao-Bao Yang}
\thanks{Corresponding author: scxbyang@scut.edu.cn}
\address {Department of Physics, South China University of Technology, Guangzhou, Guangdong 510640, China}

\begin{abstract}
Low-temperature expansion of Ising model has long been a topic of significant interest in condensed matter and statistical physics. In this paper we present new results of the coefficients in the low-temperature series of the Ising partition function on the square lattice, in the cases of a zero field and of an imaginary field $i(\pi/2)k_BT$. The coefficients in the low-temperature series of the free energy in the thermodynamic limit are represented using the explicit expression of the density function of the Fisher zeros. The asymptotic behaviour of the sequence of the coefficients when the order goes to infinity is determined exactly, for both the series of the free energy and of the partition function. Our analytic and numerical results demonstrate that, the convergence radius of the sequence is dependent on the accumulation points of the Fisher zeros which have the smallest modulus. In the zero field case this accumulation point is the physical critical point, while in the imaginary field case it corresponds to a non-physical singularity. We further discuss the relation between the series coefficients and the energy state degeneracies, using the combinatorial expression of the coefficients and the subgraph expansion. 
\end{abstract}

\keywords{Ising model, low-temperature series expansion, Fisher zeros, energy state degeneracy}
\maketitle

\section{Introduction}
Spin model of lattice systems is one of the most important tools in studying the phase transition via statistical mechanics. The simplest and perhaps the most well-known version of these models is the Ising model, which was first introduced by Lenz and Ising in the 1920s \cite{RN75, RN441, RN312}. After the one-dimensional Ising model was exactly solved \cite{RN75}, various approximations and quantitative analysis for Ising-like models have been proposed \cite{RN462, RN461, RN460, RN463, RN236, RN237, RN464}. In 1944, Onsager derived the exact solution of the square lattice Ising model in the absence of a magnetic field \cite{RN72}. This famous result provided the first exact expression for the partition function of two-dimensional Ising models. Ever since, the square lattice Ising model has been one of the most fundamental and profound systems in statistical physics of lattice models \cite{RN465}. Later in 1952, Lee and Yang presented the expression for the square lattice Ising model in an imaginary field $i(\pi/2)k_BT$ \cite{RN57}. Solutions of the models on some other typical lattices in two dimensions, such as the honeycomb \cite{RN122, RN123}, the triangular \cite{RN81, RN220}, the Kagomé \cite{RN121, RN82} and the checkerboard lattices \cite{RN58}, were also reported. The imaginary field case of each of these models had been solved \cite{RN67, RN68, RN51, RN274, RN469}.

In this work we focus on the square lattice Ising model with isotropic interactions, considering the topic of low-temperature series expansion of the free energy and partition function. Two exactly solvable cases are studied---one is in a zero field and the other is in an imaginary field $i(\pi/2)k_BT$. For a system consisting of $N$ spins $\left\{ {s_i} =  \pm 1,~i = 1, \cdots ,N \right\}$, we denote the interactions by $J$ and the external magnetic field by $H_{\rm{ex}}$. The Hamiltonian is expressed as 
\begin{equation}
H = \sum\limits_{\left\langle {ij} \right\rangle } {J{s_i}{s_j}} - {H_{{\rm{ex}}}}\sum\limits_{i = 1}^N {{s_i}},  \label{eq1}
\end{equation}
where the sum $\sum_{\left\langle {ij} \right\rangle }$ is over all nearest-neighbours and the field $H_{\rm{ex}}$ can be 0 or $i(\pi/2)k_BT$. The partition function is then defined as the sum of the Boltzmann factors over all possible configurations
\begin{equation}
Z = \sum\limits_{\left\{ {{s_i}} \right\} =  \pm 1} {{e^{ - \beta H\left( {\left\{ {{s_i}} \right\}} \right)}}}~,  \label{eq2}
\end{equation}
with $\beta  = 1/{{k_B}T}$. In the zero field case, Onsager's famous solution in the thermodynamic limit \cite{RN72} is 
\begin{align}
\mathop {\lim }\limits_{N \to \infty } \frac{1}{N}\ln Z = \ln 2 &+ \frac{1}{8{\pi ^2}}\int_0^{2\pi } d\theta \int_0^{2\pi } d\varphi \ln \left[ {{\cosh }^2}\left( {2\beta J} \right) \right. \nonumber \\ 
& \left. + \sinh \left( {2\beta J} \right)\left( {\cos \theta  + \cos \varphi } \right) \right],   \label{eq3}
\end{align}
which has also been derived in various works \cite{RN73, RN267, RN74, RN207, RN76, RN265, RN218, RN269, RN209}. In the presence of the imaginary field, the exact expression for $Z \left( i\frac{\pi }{2} \right)$ becomes \cite{RN57, RN67, RN71, RN68, RN69, RN70, RN66, RN274}
\begin{align}
&\mathop {\lim }\limits_{N \to \infty } \frac{1}{N}\ln Z \left( i\frac{\pi }{2} \right) = i\frac{\pi }{2} + \frac{1}{16{\pi ^2}}\int_0^{2\pi } d\theta \int_0^{2\pi } d\varphi \ln \left[ 16 \right.  \nonumber \\ 
&\left. \times {{\cosh }^2}\left( {2\beta J} \right){{\sinh }^2}\left( {2\beta J} \right) - 8{{\sinh }^2}\left( 2\beta J \right) \left( {\cos \theta  + \cos \varphi } \right) \right].   \label{eq4}
\end{align}
We set $J<0$ in this paper, and the case that $J>0$ is symmetric.

The low-temperature series of square lattice Ising model had been introduced by Kramers and Wannier in 1941 \cite{RN237}. Domb suggested that sufficiently lengthy series might provide a direct assessment of critical behaviour, and derived the celebrated series for the partition function in 1949 \cite{RN290, RN451}. Later in 1960, Domb further proposed the series for the free energy \cite{RN284, RN472}. Over the past decades, the series expansion has become one of the most successful methods to elucidate the properties of Ising model. The low-temperature expansion, which is the focus of this paper, has received considerable attention \blue{\cite{RN291, RN239, RN471, RN429, RN442, RN507, RN425, RN426, RN414}}. \blue{Since the convergence radius of the series is directly determined by the dominant singularity of the partition function, understanding the exact relation between the series coefficients and the singularities is very important. Expression of the coefficients by the distribution of partition function singularities will be very helpful for our understanding.}

Our work is inspired by a remarkable work in 2016 \cite{RN414}\blue{,} in which the exact expression for the coefficients in the zero field is given in terms of Bell polynomials. It is the aim of the present paper to investigate the properties of the coefficients, in particular the exact asymptotic form when the order goes to infinity, using a different approach. We use the Fisher zeros \cite{RN308}, i.e. the partition function zeros \blue{(singularities)} in the complex temperature plane, in our analysis for both the zero field and imaginary field cases. \blue{The relation between the coefficients and the Fisher zeros, which is the key point of this paper, is clearly shown. Futhermore, we indicate that the coefficients can be naturally connected to the energy state degenaracies. To generate the low-temperature series for our numerical examination, we adopt an efficient method} which was introduced in very recent works evaluating the partition function via a hypergeometric series \cite{RN423, RN424}.

The paper is organized as follows. In Sec.~\ref{fisher zeros} we revisit the density function of the Fisher zeros of square lattice Ising model in both cases. In Sec.~\ref{zero-field} and Sec.~\ref{imaginary-field} we present the results of the coefficients in the low-temperature series for the zero field and imaginary field cases, respectively. The explicit expressions of the density function of the Fisher zeros are employed to represent the coefficients and derive the exact asymptotic form of the sequence. The relation between the coefficients and the energy state degeneracies is discussed. Summary and discussion are given in Sec.~\ref{summary}.

\section{Fisher Zeros of the Square Lattice Ising Model}  \label{fisher zeros}

\begin{figure} 
\includegraphics{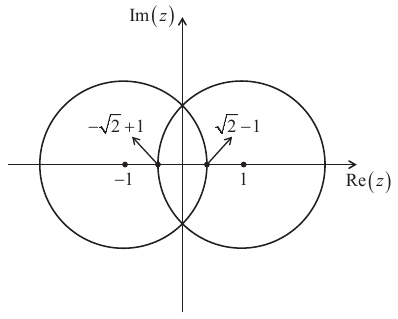}
\caption{The accumulation points of Fisher zeros in the complex $z=e^{2\beta J}$ plane in the zero field case [Eq. (\ref{eq5})].} \label{fig1}
\end{figure}

The study of partition function zeros of Ising model was initiated by the works of Lee and Yang \cite{RN303, RN57}, in which the relation between the partition function zeros and phase transition for a general system \cite{RN479} is elucidated. In the context of Ising models, Lee-Yang zeros usually refer to the partition function zeros in the complex field plane. In 1965, Fisher first considered the partition function zeros in the complex temperature plane \cite{RN308}, which had been named as ``Fisher zeros'' since then. He conjectured that such zeros for the square lattice Ising model in the absence of a magnetic field lie on two circles in the thermodynamic limit. To be concrete, in the complex $z=e^{2\beta J}$ plane the zeros approach two circles 
\begin{equation}
\left| z \pm 1 \right| = \sqrt 2    \label{eq5}
\end{equation}
in the thermodynamic limit, as shown in Fig.~\ref{fig1}. In 1974, Brascamp and Kunz introduced special boundary conditions for the square lattice Ising model \cite{RN410}, under which the solution of finite lattice can be expressed in a product form. For a square lattice of $M$ rows and $2L$ columns, the so-called Brascamp-Kunz boundary conditions are as follows: (i) In the $L$ direction the lattice is under the periodic boundary conditions. (ii) Two edges of the cylinder are fixed, with the upper edge consisting of $2L$ ``$+$'' spins and the lower edge consisting of $2L$ alternating spins ``$+-\cdots+-$''. In the absence of a magnetic field, the partition function of this case is \cite{RN410, RN455, RN475}:
\begin{equation}
Z_{2ML} = 2^{2ML}\prod\limits_{i = 1}^L \prod\limits_{j = 1}^M \left[ {1 + {\bar z}^2 + \bar z\left( {\cos {\theta _i} + \cos {\varphi _j}} \right)} \right]  \label{eq6}
\end{equation}
where $\bar z = \sinh \left( {2\beta J} \right)$, ${\theta _i} = \frac{\left( {2i - 1} \right)\pi }{2L}$ and ${\varphi _j} = \frac{j\pi }{M + 1}$. Taking the thermodynamic limit $M \to \infty,~L \to \infty$ one can easily obtain Eq.~(\ref{eq3}). This product form allows us to determine the Fisher zeros conveniently. It is obvious that, in the complex $\bar z$ plane the zeros are $\bar z = e^{\pm i\alpha _{ij}}$ with $\cos {\alpha _{ij}} = - \frac{1}{2}\left( {\cos {\theta _i} + \cos {\varphi _j}} \right)$. So that, the $2ML$ zeros lie on the unit circle $\left| \bar z \right|=1$. Then we can calculate the $4ML$ zeros $\left\{ z_i,~i = 1, \cdots ,4ML \right\}$ in the variable $z=e^{2\beta J}$ and find that they are precisely on two circles in Eq.~(\ref{eq5}), even for this finite lattice. As the system approaches the thermodynamic limit, the accumulation points of the zeros form these two circles. The number of zeros increases to infinity and the distribution on two circles can be described by a density function \cite{RN57, RN308, RN427}. 

Since the thermodynamic limit of a system without long-range interactions is independent of the boundary conditions, Lu and Wu employed the Brascamp-Kunz boundary conditions to derive the density function of Fisher zeros \cite{RN411}. They presented the explicit expressions of the densities in the complex $\bar z$ plane and $\tanh \left( - \beta J \right)$ plane, but the transformation into the variable $e^{2\beta J}$ is straightforward. For the zeros on two circles $\pm 1 + \sqrt 2 e^{i\theta } \left(0 \le \theta  < 2\pi \right)$ in Fig.~\ref{fig1}, the explicit expression of the density function is 
\begin{align}
&g_{\rm{l}}\left( \theta  \right) = \frac{{2\left| {\sin \theta } \right|\left( {\sqrt 2  - \cos \theta } \right)\left| {1 - \sqrt 2 \cos \theta } \right|}}{{{\pi ^2}{{\left( {3 - 2\sqrt 2 \cos \theta } \right)}^2}}} \times   \nonumber \\
&~~~~~~~~~~~K\left( {\frac{{2\sin \theta \left( {\sqrt 2  - \cos \theta } \right)}}{{3 - 2\sqrt 2 \cos \theta }}} \right),   \nonumber \\
&g_{\rm{r}}\left( \theta  \right) = g_{\rm{l}}\left( {\pi  - \theta } \right),  \label{eq7}
\end{align}
where $K\left( k \right) = \int_0^{{\pi  \mathord{\left/ {\vphantom {\pi  2}} \right. \kern-\nulldelimiterspace} 2}} {\frac{1}{{\sqrt {1 - {k^2}{{\sin }^2}\phi } }}d\phi }$ is the complete elliptic integral of the first kind, l and r correspond to the circles on the left and right, respectively. This density function satisfies
\begin{equation}
\int_0^{2\pi } {g_{\rm{l}}\left( \theta  \right)d\theta } = \frac{1}{2},~\int_0^{2\pi } {g_{\rm{r}}\left( \theta  \right)d\theta } = \frac{1}{2}.  \label{eq8}
\end{equation}
The plot of this density can be seen in Fig. 2(b) of Ref. \cite{RN411}. Now we show how the free energy can be expressed by this density function. Consider the low-temperature expansion of the partition function. It can be verified that there is only one configuration in the lowest energy $4MLJ$ and only one in the highest energy $-4MLJ$. The number of the first-excited energy states is $L$. Then the partition function is determined by the Fisher zeros $\left\{ z_i \right\}$:
\begin{eqnarray}
Z_{2ML} &=& {e^{ - \beta \left( {4MLJ} \right)}}\left[ 1 + L{e^{\beta \left( {4J} \right)}} +  \cdots  + {e^{\beta \left( {8MLJ} \right)}} \right]   \nonumber \\
&=& z^{- 2ML}\left[ {1 + L{z^2} +  \cdots  + {z^{4ML}}} \right]   \nonumber \\
&=& z^{- 2ML}\prod\limits_{i = 1}^{4ML} {\left( z - {z_i} \right)}.   \label{eq9}
\end{eqnarray}
Taking the thermodynamic limit leads to
\begin{align}
&~~~~\mathop {\lim }\limits_{N \to \infty } \frac{1}{N}\ln Z    \nonumber \\
&= \mathop {\lim }\limits_{M \to \infty ,L \to \infty } \frac{1}{{2ML}}\ln {Z_{2ML}}   \nonumber \\
&= - \ln z + 2\int_0^{2\pi } {g_{\rm{l}}\left( \theta  \right)\ln \left[ {z - \left( { - 1 + \sqrt 2 {e^{i\theta }}} \right)} \right]d\theta }  \nonumber \\
&~~~~+ 2\int_0^{2\pi } {g_{\rm{r}}\left( \theta  \right)\ln \left[ {z - \left( {1 + \sqrt 2 {e^{i\theta }}} \right)} \right]d\theta }.   \label{eq10}
\end{align}
Eq.~(\ref{eq3}) is now re-expressed via the density function. We note that there has been a paper arguing that the definition of the density function by Lu and Wu \cite{RN411} is not rigorous \cite{RN428}. However, in our consideration of the present study, we only need to make use of the fact that Eq.~(\ref{eq10}) is exact. Therefore, our result is not affected by the definition of the density of Fisher zeros. We also remark the special accumulation points $\pm \left( {\sqrt 2  - 1} \right)$, which have the smallest modulus and will play the key role in the analysis of the low-temperature series. The circles cut the interval $\left[ {0,1} \right]$ on the positive real axis at $z = \sqrt 2  - 1$, which is the physical critical point. As pointed out by Fisher \cite{RN308} and Lu and Wu \cite{RN411}, the density function near the critical point exhibits a linear behaviour at small $\theta$. From Eq.~(\ref{eq7}) we have 
\begin{equation}
\mathop {\lim }\limits_{\theta  \to 0} \frac{{g_{\rm{l}}\left( \theta  \right)}}{{\left| {\sin \theta } \right|}} = \frac{3 + 2\sqrt 2}{\pi }.   \label{eq11}
\end{equation}
This leads to the singularity of ${{{\partial ^2}\left( {\mathop {\lim }\limits_{N \to \infty } \frac{1}{N}\ln Z} \right)} \mathord{\left/ {\vphantom {{{\partial ^2}\left( {\mathop {\lim }\limits_{N \to \infty } \frac{1}{N}\ln Z} \right)} {\partial {z^2}}}} \right. \kern-\nulldelimiterspace} {\partial {z^2}}}$ at the critical point. Hence, there is a divergence of the specific heat, i.e. a second-order phase transition.

\begin{figure} 
\includegraphics{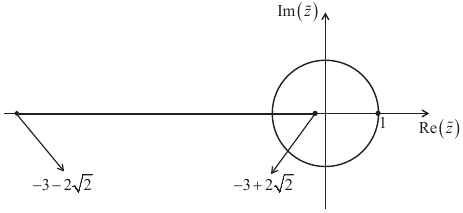}
\caption{The accumulation points of Fisher zeros in the complex $\tilde z=e^{4\beta J}$ plane in the imaginary field case [Eqs.~(\ref{eq17})-(\ref{eq18})].} \label{fig2}
\end{figure}

In an imaginary field $i(\pi/2)k_BT$, the solution of the finite lattice $M\times2L$ under the Brascamp-Kunz boundary conditions has also been obtained \cite{RN432}. When $M$ is even, the partition function is
\begin{align}
&Z_{2ML}\left( i\frac{\pi }{2} \right) = \left( {\frac{{1 - \tilde z}}{{\tilde z}}} \right)^{ML}\prod\limits_{i = 1}^L \prod\limits_{j = 1}^{M/2} \left[ 1 + {\tilde z}^2 \right.  \nonumber \\
&~~~~~~~~~~~~~~~~~~~~~~~~~~~\left.+ 2\tilde z\left( 1 - {\cos {2\tilde\theta _i} - \cos {\tilde\varphi _j}} \right) \right]   \label{eq12}
\end{align}
with $\tilde z = e^{4\beta J}$, $\tilde\theta _i = \frac{\left( 2i - 1 \right)\pi}{2L}$ and $\tilde\varphi _j = \frac{\left( 2j - 1 \right)\pi}{M + 1}$. Taking the thermodynamic limit leads to Eq.~(\ref{eq4}), except for a constant $i\frac{\pi }{2}$ which does not affect the correctness. Actually, we notice that
\begin{equation}
e^{\beta H_{\rm{ex}}\sum\limits_{i = 1}^{2ML} {s_i}} = {i^{\sum\limits_{i = 1}^{2ML} {{s_i}} }} = {i^{2ML}}\prod\limits_{i = 1}^{2ML} {{s_i}}  \label{eq13}
\end{equation}
by using the identity $i^{s_i} = i\times{s_i}$ \cite{RN51, RN274}. The constant $i\frac{\pi }{2}$ in Eq.~(\ref{eq4}) arises from the logarithm of $i^{2ML}$. In this case $2ML$ is multiples of 4 so that $i^{2ML}=1$, thus this constant can be omitted. Now we see that the effect of the field $H_{\rm{ex}}$ is formulated as the product of all spins and the partition function can be expressed as
\begin{equation}
Z\left( i\frac{\pi }{2} \right) = \sum\limits_{\left\{ {s_i} \right\} =  \pm 1} \left(\prod\limits_i {s_i}\right) e^{ - \beta \sum\limits_{\left\langle {ij} \right\rangle} J{s_i}{s_j} }.  \label{eq14}
\end{equation}
The factor $\prod\limits_i {s_i}$ is included in the contribution of a certain configuration, and the coefficients in the low-temperature series may differ from those in the zero field case. We write down the expansion by analogy with Eq.~(\ref{eq9}):
\begin{align}
Z_{2ML}\left( i\frac{\pi }{2} \right) &= {e^{ - \beta \left( {4MLJ} \right)}}\left[ 1 + \left(-L\right)e^{4\beta J } +  \cdots  + e^{8ML\beta J} \right]   \nonumber \\
&= \tilde z^{-ML}\left[ 1 + \left(-L\right)\tilde z +  \cdots  + \tilde z^{2ML} \right]   \nonumber \\
&= \tilde z^{-ML}\left( 1 - \tilde z \right)^{ML}\prod\limits_{i = 1}^{ML} {\left( \tilde z - {\tilde z_i} \right)}.   \label{eq15}
\end{align}
Here we have used the notation $\left\{ {\tilde z}_i,~i = 1, \cdots ,ML \right\}$ for the zeros except for the root 1 of multiplicity $ML$. That is,
\begin{equation}
\prod\limits_{i = 1}^{ML} {\left( \tilde z - {\tilde z_i} \right)} = \prod\limits_{i = 1}^L \prod\limits_{j = 1}^{M/2} \left[ 1 + {\tilde z}^2 + 2\tilde z\left( 1 - \cos {2\tilde\theta _i} - \cos {\tilde\varphi _j} \right) \right].     \label{eq16}
\end{equation}
As pointed out in Ref.~\cite{RN432}, the $ML$ zeros $\left\{ {\tilde z}_i \right\}$ lie on the unit circle
\begin{equation}
\left| \tilde z \right| = 1,~{\rm{for}}~-1 \le 1 - \cos {2\tilde\theta _i} - \cos {\tilde\varphi _j} \le 1,   \label{eq17}
\end{equation}
and on the line segment
\begin{equation}
-3-2\sqrt 2 \le \tilde z \le -3+2\sqrt 2 ,~{\rm{for}}~1 < 1 - \cos {2\tilde\theta _i} - \cos {\tilde\varphi _j} \le 3.   \label{eq18}
\end{equation}
In the thermodynamic limit the accumulation points of zeros form these two loci, as shown in Fig.~\ref{fig2}. The accumulation points do not cut the interval $\left[ 0,1 \right]$ on the positive real axis, therefore the system does not exhibit a physical phase transition.

Lu and Wu also found the density function of the zeros in this case \cite{RN411}. On the unit circle $e^{i\alpha}\left( {0 \le \alpha  < 2\pi } \right)$ the density is 
\begin{equation}
g_{\rm{c}} \left( \alpha \right) = \frac{\left| \sin \alpha \right|}{2{\pi ^2}}K\left( \sqrt {\frac{{\left( {3 + \cos \alpha } \right)\left( {1 - \cos \alpha } \right)}}{4}} \right),   \label{eq19}
\end{equation}
while on the line segment $- e^\lambda$ $\left( \ln \left( {3 - 2\sqrt 2 } \right) \le \lambda  \le \ln \left( {3 + 2\sqrt 2 } \right) \right)$ the density is
\begin{eqnarray}
\!\!\!\!\! g_{\rm{li}} \left( \lambda \right) = \frac{\left| \sinh \lambda \right|}{2{\pi ^2}}K\left( \sqrt {\frac{{\left( {3 - \cosh \lambda } \right)\left( {1 + \cosh \lambda } \right)}}{4}} \right).  \label{eq20}
\end{eqnarray}
Here c and li refer to the circle and line segment, respectively. The plots of Eqs.~(\ref{eq19}) and (\ref{eq20}) can be seen in Fig. 4 of Ref. \cite{RN411}. From Eq.~(\ref{eq15}) we can re-express the free energy in Eq.~(\ref{eq4}) via the density function:
\begin{align}
&~~~~\mathop {\lim }\limits_{N \to \infty } \frac{1}{N}\ln Z \left( i\frac{\pi }{2} \right)  \nonumber \\ 
&= \mathop {\lim }\limits_{M \to \infty ,L \to \infty } \frac{1}{2ML}\ln Z_{2ML}\left( i\frac{\pi }{2} \right)   \nonumber \\
&= -\frac{1}{2} \ln \tilde z + \frac{1}{2} \ln \left( 1 - \tilde z \right) + \frac{1}{2} \int_0^{2\pi } { g_{\rm{c}} \left( \alpha \right) \ln \left( \tilde z - e^{i\alpha} \right) d\alpha }  \nonumber \\
&~~~~+ \frac{1}{2}\int_{\ln \left( 3 - 2\sqrt 2 \right)}^{\ln \left( 3 + 2\sqrt 2 \right)} { g_{\rm{li}} \left( \lambda \right) \ln \left( \tilde z + e^\lambda \right) d\lambda }.  \label{eq21}
\end{align}
We remark the special accumulation point $-3+2\sqrt 2$ with the smallest modulus. At this point the density function has a finite value
\begin{equation}
g_{\rm{li}} \left( \ln \left( 3 - 2\sqrt 2 \right) \right) = \frac{1}{\sqrt 2 \pi }~.   \label{eq22}  
\end{equation}
Thus we know that this is a non-physical critical point of the first order.

\blue{When the field is neither 0 nor $i(\pi/2)k_BT$, the partition function has not been exactly solved. Thus, this case is not included in our analysis. We note that there have been numerical studies of the Fisher zeros in finite size; see for example, Ref. \cite{RN505} for real $e^{-2\beta H_{\rm{ex}}} \in \left[-1,1\right]$ and Ref. \cite{RN506} for complex $e^{-2\beta H_{\rm{ex}}} = e^{i\theta} (0<\theta<\pi)$.}

In the end of this section we emphasize that Eqs.~(\ref{eq10}) and (\ref{eq21}) are the basis of the analysis of the low-temperature series. The accumulation point with the smallest modulus in either case will play the key role in determining the asymptotic form of the sequence of the coefficients. In either case the convergence radius of the sequence is dependent on the modulus of this accumulation point.

\section{Results of the Zero-field Case}  \label{zero-field}
In Sec.~\ref{zero-field} and Sec.~\ref{imaginary-field} we consider the low-temperature series expansion of the square lattice Ising model under the periodic boundary conditions. For a finite square lattice of $N$ spins, the expansion of the partition function involving all energy states can be obtained:
\begin{align}
Z &= 2{e^{ - \beta \left( {2NJ} \right)}}\left[ 1 + N{e^{8\beta J}} + 2N{e^{12\beta J}} +  \cdots  + e^{\beta \left( {4NJ} \right)} \right]   \nonumber \\
&= 2{z^{ - N}}\left[ 1 + N{z^4} + 2N{z^6} +  \cdots  + z^{2N} \right].   \label{eq23}
\end{align}
The expansion starts from the ground state where all spins are ``$+$'' and the energy is $2NJ$. The first-excited level is $-8J$ above the ground state, as one spin changes into ``$-$'' and four pairs of interactions change from ``$+1$'' into ``$-1$''. There are $N$ configurations in this level, i.e. the number of spins. The second-excited states are those configurations with one pair of nearest-neighbour spins changing into ``$-$''. The energy is $-12J$ above the ground state with six pairs of interactions changing from ``$+1$'' into ``$-1$''. The number of states is $2N$, i.e. the number of edges in the lattice. Taking all energy levels into account, Eq.~(\ref{eq23}) can be written as
\begin{equation}
Z = 2{z^{ - N}}\sum\limits_{n = 0}^{2N} {g_n}{z^n}  \label{eq24}
\end{equation}
with 
\begin{equation}
g_0 = 1,~g_4 = N,~g_6 = 2N,~\cdots,~g_{2N-4} = N,~g_{2N} = 1   \label{eq25}
\end{equation}
and others equal to 0. The factor 2 arises from the fact that reverse of all spins conserves the energy, and $2g_n$ is the realistic degeneracy. Each term $z^n$ corresponds to the level where $n$ pairs of interactions change sign and the energy gains $-2nJ$ above the ground state. 

Now we show the celebrated series in the thermodynamic limit by Domb \cite{RN290, RN451, RN284, RN472}:
\begin{equation}
\mathop {\lim }\limits_{N \to \infty } \frac{1}{N}\ln Z =  - \ln z + \sum\limits_{n = 1}^\infty {a_n}{z^n}   \label{eq26}
\end{equation}
for the free energy, and
\begin{equation}
\mathop {\lim }\limits_{N \to \infty } Z^{{1 \mathord{\left/ {\vphantom {1 N}} \right. \kern-\nulldelimiterspace} N}} = z^{-1}\left( {1 + \sum\limits_{n = 1}^\infty  {{b_n}{z^n}} } \right)   \label{eq27}
\end{equation}
for the partition function per spin. Either expression should be regarded as a formal power series. It is clear from comparison with Eq.~(\ref{eq24}):
\begin{equation}
\sum\limits_{n = 1}^\infty  {{a_n}{z^n}}  = \mathop {\lim }\limits_{N \to \infty } \frac{1}{N}\ln \left( {\sum\limits_{n = 0}^{2N} {{g_n}{z^n}} } \right)   \label{eq28}
\end{equation}
and
\begin{equation}
1 + \sum\limits_{n = 1}^\infty  {{b_n}{z^n}}  = \mathop {\lim }\limits_{N \to \infty } {\left( {\sum\limits_{n = 0}^{2N} {{g_n}{z^n}} } \right)^{{1 \mathord{\left/ {\vphantom {1 N}} \right. \kern-\nulldelimiterspace} N}}} = \exp \left( {\sum\limits_{n = 1}^\infty  {{a_n}{z^n}} } \right).  \label{eq29}
\end{equation}     
From the exact solution in Eq.~(\ref{eq3}) we can verify that, $\sum\limits_n {a_n}{z^n}$ is actually the power series expansion of a double integral in terms of $z$: 
\begin{align}
&\sum\limits_{n = 1}^\infty {a_n}{z^n} = \frac{1}{8{\pi ^2}}\int_0^{2\pi } d\theta \int_0^{2\pi } d\varphi \ln \left[ z^4 + 2{z^2} + 1  \right.    \nonumber \\
&~~~~~~~~~~~~~~~~~~~~~~~~~~~ \left. + 2\left( z^3 - z \right)\left( \cos \theta  + \cos \varphi  \right) \right].   \label{eq30}
\end{align}

\blue{For the convenience of readers, we point out the main results in the beginning of this section. The exact asymptotic forms of $\left\{a_n\right\}$ and $\left\{b_n\right\}$ are Eq. (\ref{eq43}) and Eq. (\ref{eq56}), respectively. The relation between $\left\{a_n\right\}$, $\left\{b_n\right\}$ and the energy state degeneracies can be seen in Eqs.~(\ref{eq61})-(\ref{eq63}).}

\subsection{Asymptotic form of $\left\{a_n\right\}$}
Since $\sum\limits_n {a_n}{z^n}$ is a formal power series, it is straightforward to express the coefficients as
\begin{eqnarray*}
a_n = \frac{1}{n!} \left. \frac{\partial ^n \left( \mathop {\lim }\limits_{N \to \infty } \frac{1}{N}\ln Z + \ln z \right)}{\partial z^n} \right|_{z = 0} .
\end{eqnarray*}
Then $a_n$ can be represented by the density function of Fisher zeros from Eq.~(\ref{eq10}):
\begin{align*}
&a_n =  - \frac{2}{n}\left[ \int_0^{2\pi } g_{\rm{l}}\left( \theta \right)\frac{1}{{{{\left( { - 1 + \sqrt 2 {e^{i\theta }}} \right)}^n}}}d\theta   \right.   \nonumber \\
&~~~~~~~~~~~~~~\left. + \int_0^{2\pi } g_{\rm{r}}\left( \theta \right)\frac{1}{{{{\left( {1 + \sqrt 2 {e^{i\theta }}} \right)}^n}}}d\theta   \right]    \nonumber \\
&= - \frac{2}{n}\int_0^{2\pi } g_{\rm{l}}\left( \theta  \right)\left[ {\frac{1}{{{{\left( { - 1 + \sqrt 2 {e^{i\theta }}} \right)}^n}}} + \frac{1}{{{{\left( { 1 - \sqrt 2 {e^{ - i\theta }}} \right)}^n}}}} \right]d\theta   
\end{align*}
In the last step we have used $g_{\rm{r}}\left( \theta \right) = g_{\rm{l}}\left( \pi  - \theta \right)$. We also notice $g_{\rm{l}}\left( \theta \right) = g_{\rm{l}}\left( 2\pi  - \theta \right)$, which gives that $a_n$ is equal to 0 when $n$ is odd, and 
\begin{equation}
a_n = - \frac{4}{n}\int_0^{2\pi } g_{\rm{l}}\left( \theta  \right)\frac{1}{{{{\left( { - 1 + \sqrt 2 {e^{i\theta }}} \right)}^n}}}d\theta    \label{eq31}
\end{equation}
when $n$ is even.

\begin{figure} 
\includegraphics{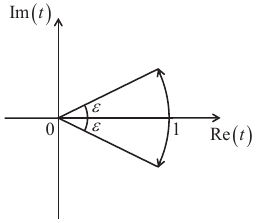}
\caption{The integral domain of the right-hand side of Eq. (\ref{eq32}).} \label{fig3}
\end{figure}

Below we give a detailed analysis for the integral in Eq.~(\ref{eq31}). It can be easily verified that the point with the largest modulus of the term $\frac{1}{{\left( - 1 + \sqrt 2 e^{i\theta} \right)}^n} \left( 0 \le \theta < 2\pi \right)$ is $\frac{1}{{\left( \sqrt 2  - 1 \right)}^n}\left( \theta  = 0 \right)$, which corresponds to the accumulation point with the smallest modulus on the circle as mentioned before. This leads to 
\begin{equation*}
\mathop {\lim }\limits_{n \to \infty } \frac{1}{{{{\left( { - 1 + \sqrt 2 {e^{i\theta }}} \right)}^n}{{\left( {\sqrt 2  + 1} \right)}^n}}} = \left\{ {\begin{array}{l}{1,~\theta  = 0}\\
{0,~\rm{otherwise}} \end{array}}. \right.  
\end{equation*}
Thus we know that when $n$ is sufficiently large the integral is mainly determined by the domain where $\left| \theta \right|$ is small. That is, the integral in $\left[-\varepsilon,\varepsilon \right]$ with a small $\varepsilon$ is of our concern:
\begin{align}
&~~~ - \int_{ - \varepsilon }^{\varepsilon}  g_{\rm{l}}\left( \theta  \right)\frac{1}{{\left( - 1 + \sqrt 2 e^{i\theta} \right)}^n}d\theta   \nonumber \\ 
& =  - \int_0^\varepsilon  \frac{g_{\rm{l}}\left( \theta  \right)}{\sin \theta } \frac{{{e^{i\theta }} - {e^{ - i\theta }}}}{{2i}}\left[ \frac{1}{{{{\left( { - 1 + \sqrt 2 {e^{i\theta }}} \right)}^n}}} \right.    \nonumber \\
&~~~~~~~~~~~~~~~~~~~~~~~~~~~~~~~~~~~\left. + \frac{1}{{{{\left( { - 1 + \sqrt 2 {e^{ - i\theta }}} \right)}^n}}} \right]d\theta   \nonumber \\
& = \frac{1}{2}\int_1^{e^{i\varepsilon }} \frac{g_{\rm{l}}\left( {\rm{Arg}}t \right)}{{\sin \left( {\rm{Arg}}t \right)}}\left[ \frac{1}{{{{\left( { - 1 + \sqrt 2 t} \right)}^n}}} + \frac{1}{{{{\left( { - 1 + \frac{{\sqrt 2 }}{t}} \right)}^n}}} \right]dt   \nonumber \\
&~~~ + \frac{1}{2}\int_1^{e^{ - i\varepsilon }} \frac{g_{\rm{l}}\left( { - {\rm{Arg}}t} \right)}{{\sin \left( { - {\rm{Arg}}t} \right)}}\left[ {\frac{1}{{{{\left( { - 1 + \frac{{\sqrt 2 }}{t}} \right)}^n}}} + \frac{1}{{{{\left( { - 1 + \sqrt 2 t} \right)}^n}}}} \right]dt.  \label{eq32}
\end{align}
The integral domain of the right-hand side of Eq.~(\ref{eq32}) is shown in Fig.~\ref{fig3}.

\begin{figure} 
\includegraphics{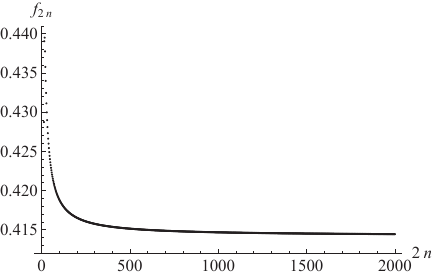}
\caption{Numerical results of $f_{2n}$ up to $2n=2000$.} \label{fig4}
\end{figure}

Then we consider 
\begin{equation*}
\int_1^{e^{i\varepsilon}} \left[ {\frac{1}{{{{\left( { - 1 + \sqrt 2 t} \right)}^n}}} + \frac{1}{{{{\left( { - 1 + \frac{{\sqrt 2 }}{t}} \right)}^n}}}} \right]dt
\end{equation*}
in the large $n$. First we have 
\begin{align}
&\int_1^{{e^{i\varepsilon }}} {\frac{1}{{{{\left( - 1 + \sqrt 2 t \right)}^n}}}dt}  = \frac{1}{\sqrt 2 \left( n - 1 \right)}\left[ \frac{1}{{{{\left( {\sqrt 2  - 1} \right)}^{n - 1}}}} \right.    \nonumber \\
&~~~~~~~~~~~~~~~~~~~~~~~~~~~~~~~~~~~~~ \left. - \frac{1}{{{{\left( {\sqrt 2 {e^{i\varepsilon }} - 1} \right)}^{n - 1}}}} \right].    \label{eq33}
\end{align}  
When $n$ is large, 
\begin{align}
\int_1^{{e^{i\varepsilon }}} {\frac{1}{{{{\left( { - 1 + \sqrt 2 t} \right)}^n}}}dt}  &\simeq \frac{1}{{\sqrt 2 \left( {n - 1} \right)}}\frac{1}{{{{\left( {\sqrt 2  - 1} \right)}^{n - 1}}}}   \nonumber \\
&= \frac{{{{\left( {\sqrt 2  + 1} \right)}^{n - 1}}}}{{\sqrt 2 \left( {n - 1} \right)}}     \label{eq34}
\end{align}
as $\left| \sqrt 2 e^{i\varepsilon } - 1 \right| > \sqrt 2  - 1$. Next we calculate
\begin{align*}
& \int_1^{{e^{i\varepsilon }}} {\frac{1}{{{{\left( { - 1 + \frac{{\sqrt 2 }}{t}} \right)}^n}}}dt} = \int_1^{{e^{i\varepsilon }}} {\frac{{{t^n}}}{{{{\left( {\sqrt 2  - t} \right)}^n}}}dt}   \nonumber \\
&~~~~~~~~~~~~~ = \int_1^{{e^{i\varepsilon }}} {{t^n}d\left[ {\frac{1}{{\left( {n - 1} \right){{\left( {\sqrt 2  - t} \right)}^{n - 1}}}}} \right]dt}  \nonumber \\
&~~~~~~~~~~~~~ = \frac{1}{n - 1}\left[ {\frac{{{e^{in\varepsilon }}}}{{{{\left( {\sqrt 2  - {e^{i\varepsilon }}} \right)}^{n - 1}}}} - \frac{1}{{{{\left( {\sqrt 2  - 1} \right)}^{n - 1}}}}} \right]   \nonumber \\
&~~~~~~~~~~~~~~~~~ - \frac{n}{n - 1}\int_1^{{e^{i\varepsilon }}} {\frac{{{t^{n - 1}}}}{{{{\left( {\sqrt 2  - t} \right)}^{n - 1}}}}dt}. 
\end{align*}
When $n$ is large, 
\begin{align*}
\int_1^{e^{i\varepsilon}} \frac{1}{{\left( - 1 + \frac{\sqrt 2}{t} \right)}^n}dt \simeq &- \frac{1}{n - 1}\frac{1}{{\left( \sqrt 2  - 1 \right)}^{n - 1}}   \nonumber \\
& - \frac{n}{n - 1}\int_1^{e^{i\varepsilon }} \frac{t^{n - 1}}{{\left( \sqrt 2  - t \right)}^{n - 1}}dt.
\end{align*}
We repeat this process, and find that for large $n$
\begin{align}
&\int_1^{e^{i\varepsilon}} \frac{1}{{{{\left( { - 1 + \frac{{\sqrt 2 }}{t}} \right)}^n}}}dt  \simeq n\left[ \frac{ - 1}{n\left( n - 1 \right)}\frac{1}{{\left( \sqrt 2 - 1 \right)}^{n - 1}} \right.    \nonumber \\
&~~~~\left. + \frac{1}{{\left( {n - 1} \right)\left( {n - 2} \right)}}\frac{1}{{{{\left( {\sqrt 2  - 1} \right)}^{n - 2}}}} +  \cdots  + \frac{ - 1}{2 \times 1}\frac{1}{\sqrt 2  - 1} \right]   \nonumber \\
&~~~~~~~~~~~~~~~~~~~~~~~~~~~ = n\sum\limits_{k = 1}^{n - 1} {\frac{{{{\left( { - 1} \right)}^k}{{\left( {\sqrt 2  + 1} \right)}^k}}}{{k\left( {k + 1} \right)}}}.  \label{eq35}
\end{align}
The sum of Eqs.~(\ref{eq34}) and (\ref{eq35}) gives that for large $n$
\begin{align}
&~~~\int_1^{e^{i\varepsilon}} \left[ \frac{1}{{\left( - 1 + \sqrt 2 t \right)}^n} + \frac{1}{{\left( - 1 + \frac{\sqrt 2}{t} \right)}^n} \right]dt   \nonumber \\
& \simeq n\left[ \frac{{\left( \sqrt 2 + 1 \right)}^{n - 1}}{\sqrt 2 n\left( n - 1 \right)} + \sum\limits_{k = 1}^{n - 1} \frac{{{\left( - 1 \right)}^k}{{\left( \sqrt 2  + 1 \right)}^k}}{k\left( k + 1 \right)} \right]   \nonumber \\
& \equiv \small{\fbox{1}}~.  \label{eq36}
\end{align}
It can be easily verified that the integral $\int_1^{e^{-i\varepsilon}}$ produces the same result
\begin{equation}
\int_1^{e^{-i\varepsilon}} \left[ \frac{1}{{\left( - 1 + \frac{\sqrt 2 }{t} \right)}^n} + \frac{1}{{\left( - 1 + \sqrt 2 t \right)}^n} \right]dt \simeq \small{\fbox{1}}~.  \label{eq37} 
\end{equation}
Obviously Eqs.~(\ref{eq36}) and (\ref{eq37}) diverge as $n \to \infty$. But when we consider
\begin{align}
&f_n = \frac{{n\left( {n - 1} \right)}}{{{{\left( {\sqrt 2  + 1} \right)}^{n - 1}}}} \times \small{\fbox{1}}   \nonumber \\
&~~= n\left[ {\frac{1}{\sqrt 2} + \frac{n\left( n - 1 \right)}{{\left( \sqrt 2 + 1 \right)}^{n - 1}}\sum\limits_{k = 1}^{n - 1} {\frac{{{\left( - 1 \right)}^k}{{\left( \sqrt 2 + 1 \right)}^k}}{{k\left( {k + 1} \right)}}} } \right]   \label{eq38}
\end{align}
we find that the sequence $\left\{ f_n \right\}$ ($n$ is even) has a limit as $n \to \infty$. Numerical results in Fig.~\ref{fig4} show that for even $n$ $\mathop {\lim }\limits_{n \to \infty } f_n \simeq 0.414 $, which we denote by $f$. Now we can see that
\begin{align}
&\mathop {\lim }\limits_{n \to \infty } \int_1^{e^{i\varepsilon}} \left[ \frac{1}{{\left( - 1 + \sqrt 2 t \right)}^n} + \frac{1}{{\left( - 1 + \frac{\sqrt 2}{t} \right)}^n} \right] \frac{n\left( n - 1 \right)}{{\left( \sqrt 2 + 1 \right)}^{n - 1}}   \nonumber \\
&~~~~~~~~~~~~~~~~~~~~~~~~~~~~~~~~~~~~~~~~~~~~~~~~~~~~~~~~~~~~~~ \times \frac{1}{f_n}dt   \nonumber \\
&=\mathop {\lim }\limits_{n \to \infty } \int_1^{e^{-i\varepsilon}} \left[ \frac{1}{{\left( - 1 + \frac{\sqrt 2}{t} \right)}^n} + \frac{1}{{\left( - 1 + \sqrt 2 t \right)}^n} \right]   \nonumber \\
&~~~~~~~~~~~~~~~~~~~~~~~~~~~~~~~~~~~~~ \times \frac{n\left( n - 1 \right)}{{\left( \sqrt 2 + 1 \right)}^{n - 1}} \frac{1}{f_n}dt   \nonumber \\
& = 1   \label{eq39}
\end{align}
and
\begin{align}
&\mathop {\lim }\limits_{n \to \infty } \left[ \frac{1}{{\left( - 1 + \sqrt 2 t \right)}^n} + \frac{1}{{\left( - 1 + \frac{\sqrt 2}{t} \right)}^n} \right] \frac{n\left( n - 1 \right)}{{\left( \sqrt 2 + 1 \right)}^{n - 1}}\frac{1}{f_n}  \nonumber \\
&~~~~~~~~~~~~~~~~~~~~~~~~~~~~~~~~~~~~ = \left\{ \begin{array}{l}{\infty,~t = 1}\\{0,~{\rm{otherwise}}}\end{array} \right. .  \label{eq40}
\end{align}
Eqs.~(\ref{eq39}) and (\ref{eq40}) leads to
\begin{align}
&\mathop {\lim }\limits_{n \to \infty } \left[ \frac{1}{{\left( - 1 + \sqrt 2 t \right)}^n} + \frac{1}{{\left( - 1 + \frac{\sqrt 2}{t} \right)}^n} \right] \frac{n\left( n - 1 \right)}{{\left( \sqrt 2 + 1 \right)}^{n - 1}}\frac{1}{f_n}  \nonumber \\
&~~~~~~~~~~~~~~~~~~~~~~~~~~~~~~~~~~~~ =\delta \left( t - 1 \right)  \label{eq41}
\end{align}
in the domain defined in Fig.~\ref{fig3}. Substituting Eq.~(\ref{eq41}) into Eqs.~(\ref{eq31}) and (\ref{eq32}) we have
\begin{align}
&~~~\mathop {\lim }\limits_{n \to \infty} a_n \times \frac{n}{4} \times \frac{n\left( n - 1 \right)}{{\left( \sqrt 2 + 1 \right)}^{n - 1}}\frac{1}{f_n}   \nonumber \\
&= \frac{1}{2}\int_1^{e^{i\varepsilon }} \frac{g_{\rm{l}} \left( {\rm{Arg}}t \right)}{\sin \left( {\rm{Arg}}t \right)} \delta \left( t - 1 \right) dt + \frac{1}{2}\int_1^{e^{ - i\varepsilon }} \frac{g_{\rm{l}} \left( { - {\rm{Arg}}t} \right)}{\sin \left( - {\rm{Arg}}t \right)}  \nonumber \\ 
&~~~~~~~~~~~~~~~~~~~~~~~~~~~~~~~~~~~~~~~~~~~~~~~~~~~~\times \delta \left( t - 1 \right) dt   \nonumber \\ 
&= \mathop {\lim }\limits_{\theta  \to 0^+ } \frac{g_{\rm{l}} \left( \theta  \right)}{ \sin \theta }   \nonumber \\
&= \frac{3 + 2\sqrt 2}{\pi }~.  \label{eq42}
\end{align}
The last step has been shown in Eq.~(\ref{eq11}). Now we obtain the exact asymptotic form of the sequence $\left\{ a_n \right\}$:
\begin{equation}
\mathop {\lim }\limits_{n \to \infty } {a_n} \times \frac{n^2\left( n - 1 \right)\pi} {4{\left( \sqrt 2 + 1 \right)}^{n + 1}f} = 1,~{\rm{for~even}}~n.     \label{eq43}
\end{equation}
When $n \to \infty$, $a_n$ is of the order $O\left( \frac{{\left( \sqrt 2 + 1 \right)}^n}{n^3} \right)$. Clearly, the convergence radius of $\sum\limits_n {a_n}{z^n}$ with respect to $z^2$ is
\begin{equation}
\mathop {\lim }\limits_{n \to \infty } \left| \frac{a_n}{a_{n + 2}} \right| = \left( {\sqrt 2  - 1} \right)^2,   \label{eq44}
\end{equation}
which exactly corresponds to the smallest modulus of the accumulation points. We note that, there has been a paper \cite{RN458} analyzing the asymptotic behaviour of high-temperature expansion coefficient for several two-dimensional Ising models. In that paper, the authors also made use of the Fisher zeros with the smallest modulus to derive the asymptotic form, but not by means of the density function. Thus, our approach differs from their method and can also be applied to the high-temperature series.

\subsection{Combinatorial expression of $\left\{a_n\right\}$ and asymptotic form of $\left\{b_n\right\}$}
The coefficient $a_n$ had been given by a combinatorial expression, using the logarithmic-polynomial expansion of the right-hand side of Eq.~(\ref{eq30}) \cite{RN414}. We quote the expression:
\begin{align}
&a_n = \frac{1}{2}\sum\limits_{\left\{ {d_1},{d_2},{d_3},{d_4} \right\}} \frac{{\left( -1 \right)}^{d_2 + d_3 + d_4 - 1}{2^{d_2}}}{d_1 + d_2 + d_3 + d_4} \frac{\left( d_1 + d_2 + d_3 + d_4 \right)!}{{d_1}!{d_2}!{d_3}!{d_4}!}   \nonumber \\
&~~~~~~~~~~~~~~~~~~~~~~~~~~~~~~~~~~~~~~~~~\times \frac{{\left[ \left( d_1 + d_3 \right)! \right]}^2}{{\left[ \left( \frac{d_1 + d_3}{2} \right)! \right]}^4},   \label{eq45}
\end{align}
where the sum is taken over all non-negative integers $\left\{ d_1, d_2, d_3, d_4 \right\}$ satisfying $d_1 + 2d_2 + 3d_3 + 4d_4 = n$ and $d_1 + d_3$ is even. One can numerically examine that this expression produces the same value as Eq.~(\ref{eq31}) for even $n$. The relation between $\left\{ b_n \right\}$ and $\left\{ a_n \right\}$ is exponential, as shown in Eq.~(\ref{eq29}). Thus $b_n$ can be expressed by the complete Bell polynomial $B_n\left( \left\{ a_j \times j! \right\} \right)$ as \cite{RN426, RN414}
\begin{equation}
b_n = \frac{1}{n!} B_n\left( \left\{ a_j \times j! \right\} \right) = \frac{1}{n!}\sum\limits_{k = 1}^n B_{n,k} \left( \left\{ a_j \times j! \right\} \right).   \label{eq46}
\end{equation}
Here $B_{n,k}\left( \left\{ a_j \times j! \right\} \right)$ represents the incomplete Bell polynomial, given by 
\begin{equation*}
B_{n,k}\left( \left\{ x_j \right\} \right) = n!\sum\limits_{\left\{ c_j \right\}} \prod\limits_{j = 1}^{n - k + 1} \frac{1}{{c_j}!}{\left( \frac{x_j}{j!} \right)}^{c_j},   
\end{equation*}
where the sum is taken over all sequences $\left\{ c_j,~j = 1, \cdots ,n-k+1 \right\}$ of non-negative integers satisfying
\begin{equation}
\sum\limits_{j = 1}^{n - k + 1} c_j = k~~{\rm{and}}~~\sum\limits_{j = 1}^{n - k + 1} j{c_j} = n.   \label{eq47}
\end{equation}
Then, $b_n$ is equal to
\begin{equation}
b_n = \sum\limits_{k = 1}^n \sum\limits_{\left\{ c_j \right\}} \prod\limits_{j = 1}^{n - k + 1} \frac{1}{{c_j}!} a_j^{c_j}   \label{eq48}
\end{equation}
with the sum $\sum\nolimits_{\left\{ c_j \right\}}{}$ taken according to Eq.~(\ref{eq47}). Or more simply, $b_n$ can be expressed as
\begin{subequations}
\begin{equation}
b_n = \sum\limits_{\left\{ \sum\limits_{j = 1}^n {j{c_j}}  = n \right\}} \prod\limits_{j = 1}^n \frac{1}{{c_j}!} a_j^{c_j}.     \label{eq49a}
\end{equation}
That is, 
\begin{equation}
b_n = a_n + \left( {a_2}{a_{n - 2}} + {a_4}{a_{n - 4}} +  \cdots  + \frac{1}{2}a_{{n \mathord{\left/ {\vphantom {n 2}} \right. \kern-\nulldelimiterspace} 2}}^2 \right) + \cdots ~.   \label{eq49b}
\end{equation}
\end{subequations}
It is straightforward to verify that $b_n$ is 0 for odd $n$, and the convergence radius of $\sum_n {b_n}{z^n}$ with respect to $z^2$ is the same as Eq.~(\ref{eq44}).

Instead of calculating $a_n$ and $b_n$ from Eqs.~(\ref{eq45}) and (\ref{eq48}), we used a very efficient method to generate these two sequences, which was introduced in very recent works \cite{RN423, RN424}. Ref. \cite{RN423} presented a hypergeometric series expansion of the free energy [Eq. (\ref{eq3})]:
\begin{align}
\mathop {\lim }\limits_{N \to \infty } \frac{1}{N}\ln Z = \ln \left[2\cosh \left( 2\beta J \right) \right] - {\kappa ^2}{}_4{F_3}\left[ \begin{array}{c} {1,1,\frac{3}{2},\frac{3}{2}}\\ 
{2,2,2} \end{array};16{\kappa ^2} \right],    \label{eq50}
\end{align}
where $\kappa  = \frac{- \tanh \left( 2\beta J \right)}{2\cosh \left( 2\beta J \right)}$ and ${}_4{F_3}\left[\cdots \right]$ represents the generalized hypergeometric series. The key finding is the proof for the hypergeometric series expansion of the double integral
\begin{align}
&~~~~\frac{1}{8{\pi ^2}}\int_0^{2\pi } d\theta \int_0^{2\pi } d\phi \ln \left[ 1 - 2\kappa \left( \cos \theta + \cos \phi \right) \right]   \nonumber \\
&=-{\kappa ^2}{}_4{F_3}\left[ \begin{array}{c} {1,1,\frac{3}{2},\frac{3}{2}}\\ {2,2,2} \end{array};16{\kappa ^2} \right],   \label{eq51}
\end{align}
which is an extension of Eq.~(109c) of Onsager's 1944 paper \cite{RN72}. Eq.~(\ref{eq50}) leads to the identities
\begin{align}
&\sum\limits_{n = 1}^\infty {a_n}{z^n} = \ln \left( {z^2} + 1 \right) - \frac{z^2 {\left( {z^2} - 1 \right)}^2}{{\left( {z^2} + 1 \right)}^4}   \nonumber \\
&~~~~~~~~~~~~~~~~~~~\times {}_4{F_3} \left[ \begin{array}{c} {1,1,\frac{3}{2},\frac{3}{2}}\\   
{2,2,2} \end{array};\frac{16{z^2}{\left( {z^2} - 1 \right)}^2}{{\left( {z^2} + 1 \right)}^4} \right]   \label{eq52}
\end{align}
(see Eq. (57) of Ref. \cite{RN424}) and
\begin{align}
&1 + \sum\limits_{n = 1}^\infty {b_n}{z^n} = \left( {z^2} + 1 \right) \exp \left\{ - \frac{z^2 {\left( {z^2} - 1 \right)}^2}{{\left( {z^2} + 1 \right)}^4} \right.   \nonumber \\
&~~~~~~~~~~~~~~~~~\left. \times {}_4{F_3} \left[ \begin{array}{c} {1,1,\frac{3}{2},\frac{3}{2}}\\   
{2,2,2} \end{array};\frac{16{z^2}{\left( {z^2} - 1 \right)}^2}{{\left( {z^2} + 1 \right)}^4} \right] \right\} .   \label{eq53}
\end{align}
Ref. \cite{RN424} used Eq.~(\ref{eq53}) and the {\sc mathematica} implementation of the ${}_p{F_q}$ hypergeometric function to generate the sequence $\left\{b_n\right\}$. The {\sc mathematica} code directly extracting the coefficients from the right-hand side of Eq.~(\ref{eq53}) allows a very fast calculation (see the code (* GM Viswanathan 2021 *) in page 12 of Ref. \cite{RN424}), and we also use this method to generate $\left\{a_n\right\}$ via Eq.~(\ref{eq52}). Below we show our results
\begin{align}
&~~~\left\{ a_2, a_4 , a_6, a_8, a_{10}, a_{12}, a_{14}, \cdots \right\}  \nonumber \\
& = \left\{ 0, 1, 2, \frac{9}{2}, 12, \frac{112}{3}, 130, \cdots \right\},  \label{eq54}
\end{align}
and 
\begin{align}
\left\{ b_2, b_4 , b_6, b_8, b_{10}, b_{12}, b_{14}, \cdots \right\} = \left\{ 0,1,2,5,14,44,152, \cdots \right\}.  \label{eq55}
\end{align}
$\left\{b_n\right\}$ is an integer sequence, and has been catalogued in the On-Line Encyclopedia of Integer Sequences (OEIS) under the number A002890 \cite{RN473}.

\begin{figure} 
\includegraphics{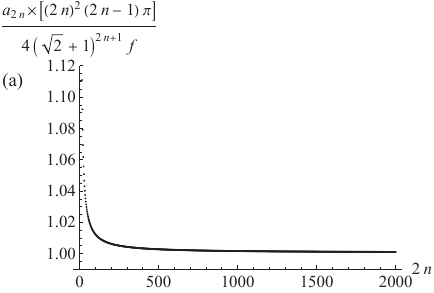} \\
\quad \\
\includegraphics{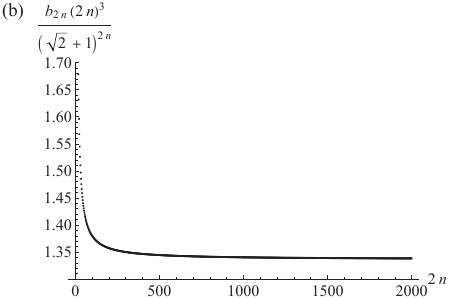}
\caption{The asymptotic forms of the low-temperature series coefficients in the zero field case. (a) Values of Eq. (\ref{eq43}) for $a_{2n}$ up to $2n=2000$. (b) Values of Eq. (\ref{eq56}) for $b_{2n}$ up to $2n=2000$.} \label{fig5}
\end{figure}

We observe that $a_n$ takes up a main part in Eq.~(\ref{eq49b}), by comparing the results in Eqs.~(\ref{eq54}) and (\ref{eq55}). Thus, we suggest that the order of $b_n$ as $n \to \infty$ is the same as $a_n$. That is, 
\begin{equation*}
\mathop {\lim }\limits_{n \to \infty } {b_n} \times \frac{n^3}{{\left( \sqrt 2 + 1 \right)}^n} = {\rm{constant}}.     
\end{equation*}
We take a numerical examination of this suggestion, shown in Fig.~\ref{fig5}(b). The numerical results confirm this suggestion, and we obtain the asymptotic form
\begin{equation}
\mathop {\lim }\limits_{n \to \infty } {b_n} \times \frac{n^3}{{\left( \sqrt 2 + 1 \right)}^n} \simeq 1.338,~{\rm{for~even}}~n.   \label{eq56}
\end{equation}
Fig.~\ref{fig5}(a) shows the asymptotic form of $\left\{ a_n \right\}$, i.e. Eq.~(\ref{eq43}).    

\subsection{Relation with the energy state degeneracies}
In Eqs.~(\ref{eq24}) and (\ref{eq25}) the realistic energy state degeneracy $2g_n$ has been introduced, in the expansion of partition function in terms of energy levels. It is obvious that the coefficients $g_n~(n \ge 0)$ are symmetric, i.e. $g_n=g_{2N-n}$. There have been many research works studying the topic of realistic degeneracy \cite{RN419, RN415, RN416, RN421, RN420, RN417, RN418}, both theoretical and numerical. As the system approaches the thermodynamic limit, the infinite low-temperature series are connected with the degeneracies, as shown by Eqs.~(\ref{eq28}) and (\ref{eq29}).

\begin{figure} 
\includegraphics{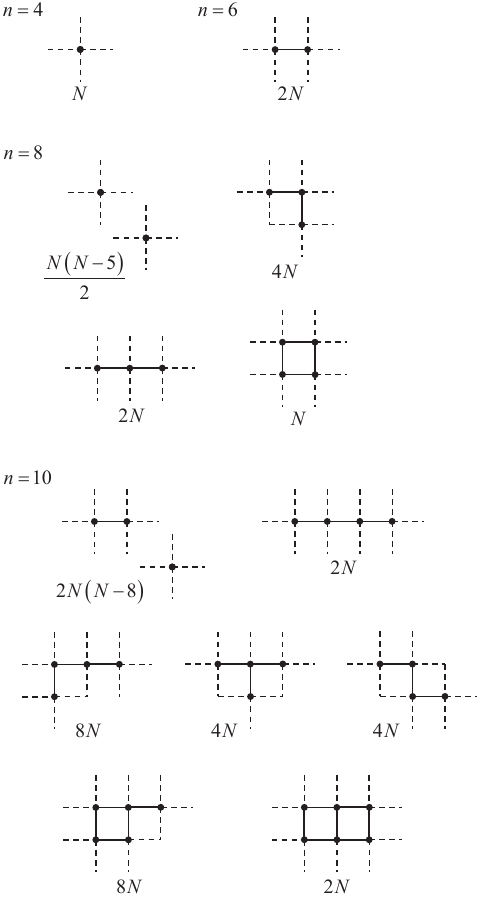}
\caption{Subgraphs contributing to $\mu_n \left( N \right)$ up to $n=10$. Dash lines represent $n$ pairs of interactions changing from ``$+1$'' to ``$-1$''. The number of configurations of each term is shown.} \label{fig6}
\end{figure}

Now we turn to consider the degeneracy in an infinite square lattice. We still start with the ground state where all spins are ``$+$''. As we have explained, the degeneracy of the level with $-2nJ$ above the ground state is the number of configurations where $n$ pairs of interactions changing from ``$+1$'' to ``$-1$''. This degeneracy in the infinite lattice should be a function of $N$, and we denote it by $2 \mu_n \left( N \right)$. We remark that the number $N$ in $\mu_n \left( N \right)$ should be regarded just as a formal variable. The coefficient $\mu_n \left( N \right)$ can be written as
\begin{equation}
\mu _n \left( N \right) = \sum\limits_{k = 1}^n \mu _{n,k}{N^k}.   \label{eq57}
\end{equation}
Fig.~\ref{fig6} displays the configurations (subgraphs) contributing to $\mu_n \left( N \right)$ up to $n=10$. Then we can obtain
\begin{align}
&\mu _0 \left( N \right) = 1,~ \mu _4 \left( N \right) = N,~ \mu _6 \left( N \right) = 2N,    \nonumber \\
&\mu _8 \left( N \right) = \frac{1}{2}{N^2} + \frac{9}{2}N,~ \mu _{10} \left( N \right) = 2{N^2} + 12N,~\cdots   \label{eq58}
\end{align}
The subgraph expansion considered here can be related to the Ursell-Mayer cluster expansion \cite{RN452, RN453, RN454}. The difference between $\mu_n$ and $g_n$ arises from the fact that, the subgraph expansion for $\mu_n$ is complete in an infinite lattice, while that for $g_n$ may be restricted by the boundary due to the finite size. For example, in a $4\times4$ lattice, the periodic boundary conditions cause that the subgraph ~$\mathop{\!\!\bullet\!\! \frac{~~}{~~} \!\!\bullet\!\! \frac{~~}{~~} \!\!\bullet\!\! \frac{~~}{~~} \!\!\bullet\!\!}$~ does not contribute to $\mu_{10}$ but to $\mu_8$. Therefore we can see that $g_n = \mu_n$ only when $n \ll N$, that is, the order $n$ is sufficiently small such that the contributing subgraphs are all within the boundary. This point has also been discussed in Ref. \cite{RN414} [see Eqs.~(10)-(12) therein]. 

In the thermodynamic limit, the exact solution can be re-expressed by
\begin{align*}
&\mathop {\lim }\limits_{N \to \infty } \frac{1}{N}\ln Z = -\ln z + \mathop {\lim }\limits_{N \to \infty } \frac{1}{N}\ln \left[ \sum\limits_{n = 0}^\infty  {\mu _n}\left( N \right){z^n} \right],   \nonumber \\
&\mathop {\lim }\limits_{N \to \infty } Z^{1 \mathord{\left/ {\vphantom {1 N}} \right. \kern-\nulldelimiterspace} N} = z^{-1}\mathop {\lim }\limits_{N \to \infty } {\left[ \sum\limits_{n = 0}^\infty  {\mu _n}\left( N \right){z^n} \right]}^{1 \mathord{\left/ {\vphantom {1 N}} \right. \kern-\nulldelimiterspace} N}.
\end{align*}
Then the low-temperature series can be written as
\begin{subequations}
\begin{equation}
\sum\limits_{n = 1}^\infty {a_n}{z^n} = \mathop {\lim }\limits_{N \to \infty } \frac{1}{N}\ln \left[ \sum\limits_{n = 0}^{\infty} \mu _{n} \left( N \right){z^n} \right]   \label{eq59a}
\end{equation}
and
\begin{equation}
1 + \sum\limits_{n = 1}^\infty {b_n}{z^n} = \mathop {\lim }\limits_{N \to \infty } {\left[ \sum\limits_{n = 0}^\infty  {\mu _n}\left( N \right){z^n} \right]}^{1 \mathord{\left/ {\vphantom {1 N}} \right. \kern-\nulldelimiterspace} N}.    \label{eq59b}
\end{equation}
\end{subequations}
Now let us consider $\exp \left(N \sum\limits_{n = 1}^\infty {a_n}{z^n} \right)$. Since $N$ serves as a formal variable here, it is natural to suggest that 
\begin{equation}
\exp \left(N \sum\limits_{n = 1}^\infty {a_n}{z^n} \right) = {\left(1 + \sum\limits_{n = 1}^\infty {b_n}{z^n}\right)}^N = \sum\limits_{n = 0}^{\infty} \mu _{n} \left( N \right){z^n}   \label{eq60}
\end{equation}
is exact for any $N$. Following Eq.~(\ref{eq60}), the coefficient $\mu _{n} \left( N \right)$ is easy to obtain by analogy with Eqs.~(\ref{eq46}) and (\ref{eq48}):
\begin{align}
&\mu _n \left( N \right) = \frac{1}{n!} B_n\left( \left\{ Na_j \times j! \right\} \right) = \sum\limits_{k = 1}^n {N^k}\sum\limits_{\left\{ c_j \right\}} \prod\limits_{j = 1}^{n - k + 1} \frac{1}{{c_j}!}a_j^{c_j}    \label{eq61}
\end{align}
with the sum $\sum\nolimits_{\left\{ c_j \right\}}{}$ taken according to Eq.~(\ref{eq47}). Comparing with Eq.~(\ref{eq57}) we know that 
\begin{equation}
\mu _{n,k} = \sum\limits_{\left\{ c_j \right\}} \prod\limits_{j = 1}^{n - k + 1} \frac{1}{{c_j}!}a_j^{c_j}.  \label{eq62}
\end{equation}
It is clear to show
\begin{equation}
a_n = \mu _{n,1},~b_n = \sum\limits_{k = 1}^n \mu _{n,k} = \mu_n \left(N=1\right).   \label{eq63}
\end{equation}
That is, $a_n$ is the coefficient associated with $N^1$ in the expression of $\mu_n \left(N\right)$ in powers of $N$, and $b_n$ is the sum of all coefficients. Being the number of contributing subgraphs of certain order, $\mu_n \left(N\right)$ should be an integer for any $N$. Thus we can see that $b_n=\mu_n \left(N=1\right)$ is always an integer. The relation between $\left\{a_n\right\}$, $\left\{b_n\right\}$ and $\left\{ \mu_n \left(N\right) \right\}$ had been pointed out by Domb in Ref. \cite{RN284} (see Sec. 3.6.1 therein). Nagle's famous work of the residual entropy of ice models \cite{RN46} had also used the relation between $\left\{b_n\right\}$ and $\left\{ \mu_n \left(N\right) \right\}$, although he was dealing with a different issue [see Eqs.~(6) and (7) therein]. 

We also examine Eq.~(\ref{eq60}). From Eqs.~(\ref{eq52}) and (\ref{eq53}) it is easy to show that the series $\sum\limits_n \mu _n \left( N \right){z^n}$ can also be evaluated via the hypergeometric function
\begin{align}
&\sum\limits_{n = 0}^{\infty} \mu _{n} \left( N \right){z^n} = \left( z^2 + 1 \right)^N \exp \left\{ - N\frac{z^2 {\left( {z^2} - 1 \right)}^2}{{\left( {z^2} + 1 \right)}^4} \right.   \nonumber \\
&~~~~~~~~~~~~~~~~~\left. \times {}_4{F_3} \left[ \begin{array}{c} {1,1,\frac{3}{2},\frac{3}{2}}\\   
{2,2,2} \end{array};\frac{16{z^2}{\left( {z^2} - 1 \right)}^2}{{\left( {z^2} + 1 \right)}^4} \right] \right\}.   \label{eq64}
\end{align}
Therefore, the method for generating $\left\{a_n\right\}$ and $\left\{b_n\right\}$ using {\sc mathematica} can be conveniently applied to $\left\{ \mu_n \left(N\right) \right\}$. We list the first terms 
\begin{align}
&\left\{ {\mu _2},{\mu _4},{\mu _6},{\mu _8},{\mu _{10}},{\mu _{12}},{\mu _{14}}, \cdots \right\} = \left\{ 0,N,2N,\frac{9N}{2} + \frac{N^2}{2}, \right.   \nonumber \\
&\left. 12N + 2{N^2},\frac{112N}{3} + \frac{13{N^2}}{2} + \frac{N^3}{6},130N + 21{N^2} + {N^3}, \cdots \right\}.    \label{eq65}
\end{align} 
Clearly, the results are exactly consistent with the subgraph expansion (see Fig.~\ref{fig6}). It is straightforward to recognize the relation in Eq.~(\ref{eq63}) by comparing Eqs.~(\ref{eq54}) and (\ref{eq55}) with Eq.~(\ref{eq65}). Hence, we have succeeded in constructing the combinatorial formula [Eq.~(\ref{eq61})] for the degeneracy in the infinite lattice, by connecting the low-temperature series coefficients with the subgraph expansion.  

\section{Results of the Imaginary-field Case}   \label{imaginary-field}
In the presence of an imaginary field $i(\pi/2)k_BT$, the product of all spins is included in the contribution of a certain configuration to the partition function, as shown in Eq.~(\ref{eq14}). In this case, the partition function involving all energy states [see Eq.~(\ref{eq23}) for the zero field case] should be 
\begin{align}
&Z\left( i\frac{\pi }{2} \right) = 2 e^{ - \beta \left( {2NJ} \right)} \left[ 1 + (-N){e^{8\beta J}} + 2N{e^{12\beta J}} +  \cdots \right.   \nonumber \\
&~~~~~~~~~~~~~~~~~~~~~~~~~~~~\left. + e^{\beta \left( {4NJ} \right)} \right]   \nonumber \\
&~~~~~~ = 2 {\tilde z}^{-{N \mathord{\left/ {\vphantom {N 2}} \right. \kern-\nulldelimiterspace} 2}} \left[ 1 + (-N){\tilde z^2} + 2N{\tilde z^3} +  \cdots + {\tilde z}^N \right].   \label{eq66}
\end{align}
Here we ignore the factor $i^N$. The low-temperature series are thereby defined as
\begin{equation}
\sum\limits_{n = 1}^\infty {\tilde a_{2n}}{\tilde z^n} = \mathop {\lim }\limits_{N \to \infty } \frac{1}{N}\ln Z\left( i\frac{\pi }{2} \right) + \frac{1}{2}\ln {\tilde z}   \label{eq67}
\end{equation}
and
\begin{equation}
1 + \sum\limits_{n = 1}^\infty  {\tilde b_{2n}}{\tilde z^n} = \mathop {\lim }\limits_{N \to \infty } Z\left( i\frac{\pi }{2} \right)^{1 \mathord{\left/ {\vphantom {1 N}} \right. \kern-\nulldelimiterspace} N} \times {\tilde z}^{1 \mathord{\left/ {\vphantom {1 2}} \right. \kern-\nulldelimiterspace} 2} = \exp\left( \sum\limits_{n = 1}^\infty {\tilde a_{2n}}{\tilde z^n} \right) .  \label{eq68}
\end{equation}
We use the notations $\tilde a_{2n}$ and $\tilde b_{2n}$ as $\tilde z = z^2$ and $\tilde z^n$ corresponds to the level where $2n$ pairs of interactions change from ``$+1$'' to ``$-1$''. From Eq. (\ref{eq4}) we see that $\sum\limits_{n} {\tilde a_{2n}}{\tilde z^n}$ is actually the power series expansion of a double integral in terms of $\tilde z$
\begin{align}
&\sum\limits_{n = 1}^\infty {\tilde a_{2n}}{\tilde z^n} = \frac{1}{16\pi^2} \int_0^{2\pi } d\theta \int_0^{2\pi } d\varphi \ln \left[ \tilde z^4 - 2{\tilde z^2} + 1  \right.    \nonumber \\
&~~~~~~~~~~~~~~~~~~ \left. - 2\left( \tilde z^3 - 2\tilde z^2 +\tilde z \right)\left( \cos \theta  + \cos \varphi  \right) \right].   \label{eq69}
\end{align}

\blue{We point out the main results in the beginning of this section. The exact asymptotic forms of $\left\{ \tilde a_{2n}\right\}$ and $\left\{ \tilde b_{2n}\right\}$ are Eq. (\ref{eq74}) and Eq. (\ref{eq84}), respectively. The relation between $\left\{ \tilde a_{2n}\right\}$, $\left\{ \tilde b_{2n}\right\}$ and the energy state degeneracies can be seen in Eqs.~(\ref{eq87})-(\ref{eq89}).}

\subsection{Asymptotic form of $\left\{\tilde a_{2n}\right\}$}
Again, we express $\tilde a_{2n}$ as
\begin{eqnarray*}
\tilde a_{2n} = \frac{1}{n!} \left. \frac{\partial ^n \left[ \mathop {\lim }\limits_{N \to \infty } \frac{1}{N}\ln Z\left( i\frac{\pi }{2} \right) + \frac{1}{2}\ln {\tilde z} \right]}{\partial {\tilde z}^n} \right|_{\tilde z = 0} .
\end{eqnarray*}
Using Eq. (\ref{eq21}) $\tilde a_{2n}$ can be represented by the density function of Fisher zeros
\begin{align}
&{\tilde a}_{2n} = -\frac{1}{2n} + \frac{-1}{2n}\int_0^{2\pi } {g_{\rm{c}} \left( \alpha \right)e^{- in\alpha} d\alpha}    \nonumber \\
&~~~~~~~~ + \frac{{\left( -1 \right)}^{n - 1}}{2n}\int_{\ln \left( 3 - 2\sqrt 2 \right)}^{\ln \left( 3 + 2\sqrt 2 \right)} {g_{\rm{li}}\left( \lambda  \right) e^{ - n\lambda}d\lambda}.     \label{eq70}
\end{align}
The derivation of the asymptotic form when $n \to \infty $ is very similar to that for $a_n$. We emphasize the accumulation point $-3+2\sqrt 2$ with the smallest modulus. When $n$ is sufficiently large, $\tilde a_{2n}$ is mainly determined by the integral on the domain $\left[ \ln \left( 3 - 2\sqrt 2 \right) ,~\ln \left( 3 - 2\sqrt 2 \right) + \gamma \right]$ with a small $\gamma$. That is,
\begin{align*}
\int_{\ln \left( 3 - 2\sqrt 2 \right)}^{\ln \left( 3 - 2\sqrt 2 \right) + \gamma} {g_{\rm{li}}\left( \lambda  \right) e^{ - n\lambda}d\lambda}~.    
\end{align*}

We consider 
\begin{equation}
\int_{\ln \left( 3 - 2\sqrt 2 \right)}^{\ln \left( 3 - 2\sqrt 2 \right) + \gamma} {{e^{ - n\lambda }}d\lambda }  = \frac{{\left( 3 + 2\sqrt 2 \right)}^n \left( 1 - e^{-n\gamma} \right)}{n}.   \label{eq71}
\end{equation}
It is easy to verify that
\begin{align}
\mathop {\lim }\limits_{n \to \infty} \frac{n e^{- n\lambda}}{{\left( 3 + 2\sqrt 2 \right)}^n \left( 1 - e^{- n\gamma} \right)} = \delta \left[ \lambda - \ln \left( 3 - 2\sqrt 2 \right) \right]    \label{eq72}
\end{align} 
in the domain $\left[ \ln \left( 3 - 2\sqrt 2 \right) ,~\ln \left( 3 - 2\sqrt 2 \right) + \gamma \right]$. Substituting Eq. (\ref{eq72}) into Eq. (\ref{eq70}) leads to 
\begin{align}
&~~~\mathop {\lim }\limits_{n \to \infty} {\tilde a}_{2n} \times \frac{ 2{n^2} {\left( -1 \right)}^{n - 1}}{{\left( 3 + 2\sqrt 2 \right)}^n}    \nonumber \\
&= \int_{\ln \left( 3 - 2\sqrt 2 \right)}^{\ln \left( 3 - 2\sqrt 2 \right) + \gamma} g_{\rm{li}} \left( \lambda  \right)\delta \left[ \lambda - \ln \left( 3 - 2\sqrt 2 \right) \right]d\lambda    \nonumber \\
&= g_{\rm{li}} \left(\ln \left( 3 - 2\sqrt 2 \right) \right)    \nonumber \\
&= \frac{1}{\sqrt 2 \pi}.    \label{eq73}
\end{align}
The last step has been shown in Eq. (\ref{eq22}). Now we obtain the exact asymptotic form of the sequence $\left\{ \tilde a_{2n} \right\}$:
\begin{equation}
\mathop {\lim }\limits_{n \to \infty} {\tilde a}_{2n} \times \frac{2\sqrt 2 \pi {n^2} {\left( -1 \right)}^{n - 1}}{{\left( \sqrt 2 + 1 \right)}^{2n}} = 1.   \label{eq74}   
\end{equation}
We can see that the series $\left\{ \tilde a_{2n} \right\}$ alternate in sign. When $n \to \infty$, $\left|\tilde a_{2n}\right|$ is of the order $O\left( \frac{{\left( \sqrt 2 + 1 \right)}^{2n}}{n^2} \right)$. The convergence radius of $\sum\limits_{n} \tilde a_{2n} \tilde z^n$ with respect to $\tilde z$ is 
\begin{equation}
\mathop {\lim }\limits_{n \to \infty} \left| \frac{\tilde a_{2n}}{\tilde a_{2n + 2}} \right| = \left( \sqrt 2 - 1 \right)^2,   \label{eq75}
\end{equation}
which exactly corresponds to the smallest modulus of the accumulation points. As mentioned in Sec. \ref{fisher zeros}, the system in the imaginary field does not have a physical phase transition. The convergence radius of the low-temperature series in this case is dependent on a non-physical singularity.

\subsection{Combinatorial expression of $\left\{ \tilde a_{2n}\right\}$ and asymptotic form of $\left\{ \tilde b_{2n}\right\}$}
Using the logarithmic-polynomial expansion of the right-hand side of Eq. (\ref{eq69}), like Ref. \cite{RN414} had done for Eq. (\ref{eq30}), we can obtain the combinatorial expression of $\tilde a_{2n}$:
\begin{align}
&\tilde a_{2n} = \frac{1}{4}\sum\limits_{\left\{ d_1,d_2,d_3,d_4,d_5 \right\}} \frac{{\left( -1 \right)}^{d_2 + d_5 - 1} 2^{d_2+d_3}}{d_1 + d_2 + d_3 + d_4 + d_5}   \nonumber \\
&~~~~\times \frac{\left( d_1 + d_2 + d_3 + d_4 + d_5 \right)!}{{d_1}!{d_2}!{d_3}!{d_4}!{d_5}!} \frac{{\left[ \left( d_1 + d_2 + d_4 \right)! \right]}^2}{{\left[ \left( \frac{d_1 + d_2 + d_4}{2} \right)! \right]}^4},   \label{eq76}
\end{align}
where the sum is taken over all non-negative integers $\left\{ d_1, d_2, d_3, d_4, d_5 \right\}$ satisfying $d_1 + 2d_2 + 2d_3 + 3d_4 + 4d_5 = n$ and $d_1 + d_2 + d_4$ is even. We do not show the details of the calculation, as the method is completely the same as that for $a_n$ in Ref. \cite{RN414} [see Eq. (\ref{eq45})]. One can numerically examine that the values produced from Eqs.~(\ref{eq70}) and (\ref{eq76}) are consistent. $\tilde b_{2n}$ is again expressed by the Bell polynomial as
\begin{equation}
\tilde b_{2n} = \frac{1}{\left( 2n \right)!}\sum\limits_{k = 1}^{2n} B_{2n,k} \left( \left\{ \tilde a_j \times j! \right\} \right) = \sum\limits_{k = 1}^{2n} \sum\limits_{\left\{ c_j \right\}} \prod\limits_{j = 1}^{2n - k + 1} \frac{1}{{c_j}!}\tilde a_j^{c_j},   \label{eq77}
\end{equation}
where $\tilde a_j=0$ when $j$ is odd, and the sum $\sum\nolimits_{\left\{ c_j \right\}}$ is taken over all sequences $\left\{ c_j,~j = 1, \cdots ,2n-k+1 \right\}$ of non-negative integers satisfying
\begin{equation}
\sum\limits_{j = 1}^{2n - k + 1} c_j = k~~{\rm{and}}~~\sum\limits_{j = 1}^{2n - k + 1} j{c_j} = 2n.   \label{eq78}
\end{equation} 
The convergence radius of $\sum\limits_{n} \tilde b_{2n} \tilde z^n$ with respect to $\tilde z$ is the same as Eq. (\ref{eq75}).

The efficient method to generate $\left\{ a_n \right\}$ and $\left\{ b_n \right\}$ using {\sc mathematica} is also straightforward to be applied to $\left\{ \tilde a_{2n} \right\}$ and $\left\{ \tilde b_{2n} \right\}$. Eq. (\ref{eq51}) enables us to present the hypergeometric series expansion of the free energy [Eq. (\ref{eq4})]
\begin{align}
&\mathop {\lim }\limits_{N \to \infty } \frac{1}{N}\ln Z \left( i\frac{\pi }{2} \right) = \frac{1}{2} \ln \left|2\sinh \left( 4\beta J \right) \right|     \nonumber \\
&~~~~~~~~~~~~~~~~~~~~~~~~~~~~ - \frac{1}{2}{\tilde \kappa ^2}{}_4{F_3}\left[ \begin{array}{c} {1,1,\frac{3}{2},\frac{3}{2}}\\ 
{2,2,2} \end{array};16{\tilde \kappa ^2} \right]    \label{eq79}
\end{align}
with $\tilde \kappa = \frac{1}{4 \cosh^{2}\left(2\beta J\right)}$. Then the low-temperature series are related to the hypergeometric function
\begin{align}
&\sum\limits_{n = 1}^\infty \tilde a_{2n}{\tilde z^n} = \frac{1}{2} \ln \left( 1- \tilde z^2 \right) - \frac{1}{2} \frac{\tilde z^2}{{\left( \tilde z + 1\right)}^4}   \nonumber \\
&~~~~~~~~~~~~~~~~~~~~~~~~~~\times {}_4{F_3} \left[ \begin{array}{c} {1,1,\frac{3}{2},\frac{3}{2}}\\ {2,2,2} \end{array};\frac{16 \tilde z^2}{{\left( \tilde z + 1\right)}^4} \right],    \label{eq80}
\end{align}
and
\begin{align}
&1 + \sum\limits_{n = 1}^\infty \tilde b_{2n}{\tilde z^n} = {\left( 1- \tilde z^2 \right)}^{1 \mathord{\left/ {\vphantom {1 2}} \right. \kern-\nulldelimiterspace} 2} \exp \left\{ - \frac{1}{2} \frac{\tilde z^2}{{\left( \tilde z + 1\right)}^4} \right.   \nonumber \\
&~~~~~~~~~~~~~~~~~~~~~~~~\left. \times {}_4{F_3} \left[ \begin{array}{c} {1,1,\frac{3}{2},\frac{3}{2}}\\ {2,2,2} \end{array};\frac{16 \tilde z^2}{{\left( \tilde z + 1\right)}^4} \right] \right\}.   \label{eq81}
\end{align}
We list the first terms of $\left\{ \tilde a_{2n} \right\}$ and $\left\{ \tilde b_{2n} \right\}$
\begin{align}
&~~~\left\{ \tilde a_2, \tilde a_4, \tilde a_6,\tilde a_8,\tilde a_{10},\tilde a_{12},\tilde a_{14}, \cdots \right\}   \nonumber \\
& = \left\{ 0, - 1,2, - \frac{15}{2},28, - \frac{346}{3},498, \cdots \right\}     \label{eq82}
\end{align}
and
\begin{align}
&~~~\left\{ \tilde b_2, \tilde b_4, \tilde b_6,\tilde b_8,\tilde b_{10},\tilde b_{12},\tilde b_{14}, \cdots \right\}   \nonumber \\
& = \left\{ 0, -1, 2, -7, 26, -106, 456, \cdots \right\}.     \label{eq83}
\end{align}
$\left\{ \tilde b_{2n} \right\}$ is an integer sequence with alternate signs.

\begin{figure} 
\includegraphics{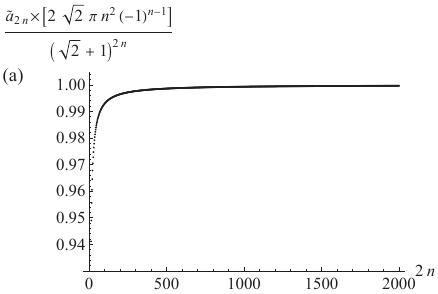} \\
\quad \\
\includegraphics{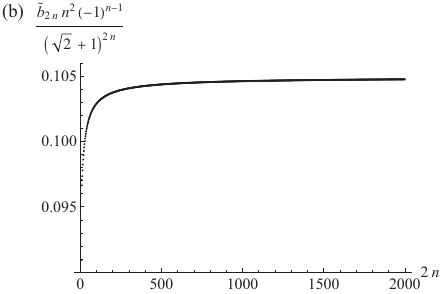}
\caption{The asymptotic forms of the low-temperature series coefficients in the imaginary field case. (a) Values of Eq. (\ref{eq74}) for $\tilde a_{2n}$ up to $2n=2000$. (b) Values of Eq. (\ref{eq84}) for $\tilde b_{2n}$ up to $2n=2000$..} \label{fig7}
\end{figure}

We still suggest that the order of $\left| \tilde b_{2n} \right|$ as $n \to \infty$ is the same as $\left| \tilde a_{2n} \right|$. The numerical results in Fig.~\ref{fig7}(b) confirm this suggestion. The asymptotic form of  $\tilde b_{2n}$ is found
\begin{equation}
\mathop {\lim }\limits_{n \to \infty } \tilde b_{2n} \times \frac{n^2 {\left(-1\right)}^{n-1}}{{\left( \sqrt 2 + 1 \right)}^{2n}} \simeq 0.105.   \label{eq84}
\end{equation}
Fig.~\ref{fig7}(a) shows the asymptotic form of $\tilde a_{2n}$, i.e. Eq. (\ref{eq74}).

\subsection{Relation with the energy state degeneracies}
\begin{figure} 
\includegraphics{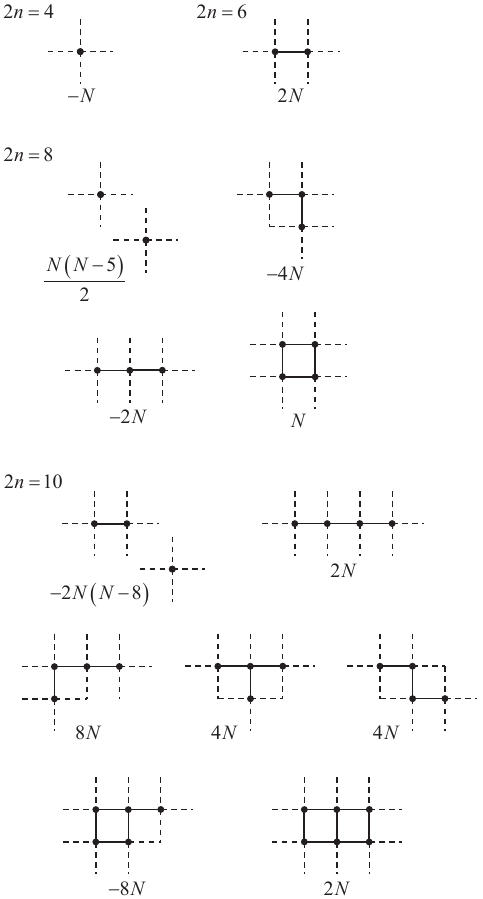}
\caption{Subgraphs contributing to $\tilde \mu_{2n} \left( N \right)$ up to $2n=10$. Dash lines represent $2n$ pairs of interactions changing from ``$+1$'' to ``$-1$''. The number of configurations of each term multiplied by ${\left( -1 \right)}^{N_{\rm{sub}}}$ is shown.} \label{fig8}
\end{figure}

The energy state degeneracy in the infinite lattice in this case can also be represented by the subgraph expansion. Since the product of all spins is taken into account, the factor ${\left( -1 \right)}^{N_{\rm{sub}}}$ should be included in the contribution with $N_{\rm{sub}}$ being the number of spins in this subgraph (the number of ``$-$'' spins of this term, note that we start with the ground state where all spins are ``$+$''). The degeneracy is denoted by $2\tilde \mu_{2n}\left( N \right)$. Fig.~\ref{fig8} displays the subgraphs contributing to $\tilde \mu_{2n}\left( N \right)$ up to $2n=10$. We see that, the difference between Figs. \ref{fig6} and \ref{fig8} is caused by ${\left( -1 \right)}^{N_{\rm{sub}}}$. From Fig.~\ref{fig8} we can obtain 
\begin{align}
&\tilde \mu _0 \left( N \right) = 1,~ \tilde \mu _4 \left( N \right) = -N,~ \tilde \mu _6 \left( N \right) = 2N,    \nonumber \\
&\tilde \mu _8 \left( N \right) = \frac{1}{2}{N^2} - \frac{15}{2}N,~ \tilde \mu _{10} \left( N \right) = -2{N^2} + 28N,~\cdots   \label{eq85}
\end{align}

By analogy with Eq. (\ref{eq60}) in the zero field case, the low-temperature series can be connected to the degeneracies by
\begin{equation}
\exp \left(N \sum\limits_{n = 1}^\infty \tilde a_{2n} \tilde z^n \right) = {\left(1 + \sum\limits_{n = 1}^\infty \tilde b_{2n} \tilde z^n \right)}^N = \sum\limits_{n = 0}^\infty \tilde \mu _{2n} \left( N \right) \tilde z^n.   \label{eq86}
\end{equation}
Again, we have constructed the combinatorial formula for $\tilde \mu_{2n}\left( N \right)$
\begin{align}
\tilde \mu_{2n} \left( N \right) = \sum\limits_{k = 1}^{2n} \tilde \mu_{2n,k} {N^k}    \label{eq87}
\end{align}
with
\begin{align}
\tilde \mu_{2n,k} = \sum\limits_{\left\{ c_j \right\}} \prod\limits_{j = 1}^{2n - k + 1} \frac{1}{{c_j}!} \tilde a_j^{c_j},    \label{eq88}
\end{align}
where the sum $\sum\nolimits_{\left\{ c_j \right\}}{}$ is taken according to Eq.~(\ref{eq78}). The relation between the coefficients is still the same as Eq. (\ref{eq63})
\begin{equation}
\tilde a_{2n} = \tilde \mu_{2n,1},~\tilde b_{2n} = \sum\limits_{k = 1}^{2n} \tilde \mu_{2n,k} = \tilde \mu_{2n} \left(N=1\right).   \label{eq89}
\end{equation}

From Eqs.~(\ref{eq80}) and (\ref{eq81}) it is straightforward to represent $\sum\limits_{n} \tilde \mu_{2n}\left(N\right) \tilde z^n$ via the hypergeometric function
\begin{align}
&\sum\limits_{n = 0}^\infty \tilde \mu _{2n} \left( N \right) \tilde z^n = {\left( 1- \tilde z^2 \right)}^{N \mathord{\left/ {\vphantom {N 2}} \right. \kern-\nulldelimiterspace} 2} \exp \left\{ -\frac{N}{2} \frac{\tilde z^2}{{\left( \tilde z + 1\right)}^4} \right.   \nonumber \\
&~~~~~~~~~~~~~~~~~~~~~~~~\left. \times {}_4{F_3} \left[ \begin{array}{c} {1,1,\frac{3}{2},\frac{3}{2}}\\ {2,2,2} \end{array};\frac{16 \tilde z^2}{{\left( \tilde z + 1\right)}^4} \right] \right\}.   \label{eq90}
\end{align}
Again, we can use the efficient method to generate $\left\{ \tilde \mu_{2n} \left(N\right) \right\}$. We list the first terms
\begin{align}
&~~~\left\{ \tilde \mu_2, \tilde \mu_4, \tilde \mu_6, \tilde \mu_8, \tilde \mu_{10}, \tilde \mu_{12}, \tilde \mu_{14}, \cdots \right\}   \nonumber \\
& = \left\{ 0, -N, 2N, -\frac{15N}{2} + \frac{N^2}{2}, 28N - 2{N^2}, \right.   \nonumber \\
&~~~~~~\left. -\frac{346N}{3} + \frac{19{N^2}}{2} - \frac{N^3}{6}, 498N - 43{N^2} + N^3, \cdots \right\}.    \label{eq91}
\end{align}

\section{Summary and discussion}   \label{summary}
In this paper, we have shown that the accumulation points of Fisher zeros with the smallest modulus play a key role in determining the asymptotic form of low-temperature series coefficient. Two exactly solvable cases of the square lattice Ising model are studied. In both cases, the series coefficient $a_n$ of the free energy is of the asymptotic form:
\begin{equation}
\left| a_n \right| \sim \frac{1}{\left| z \right|_{\min}^n n^{\nu + 1}} \left[ g_{\rm{density}} \right]\times {\rm{constant}}.     \label{eq92}
\end{equation}
Here $\left| z \right|_{\min}$ is the smallest modulus of the accumulation points, $\nu$ is the order of this singularity, and $\left[ g_{\rm{density}} \right]$ represents the appropriate property of the density function at this point [Eqs.~(\ref{eq42}) and (\ref{eq73})]. \blue{From this expression, the convergence radius can be easily obtained and the exact relation between the coefficients and the Fisher zeros is clearly established.} It can be expected that, our method using the density function of Fisher zeros can be extended to other lattices. Furthermore, the relation between the low-temperature series coefficients and the degeneracies in the infinite lattice is illustrated. We believe this relation [cf. Eqs.~(\ref{eq61})-(\ref{eq63}) and (\ref{eq87})-(\ref{eq89})] should also \blue{be applied} to other lattices.

It is interesting to consider the three-dimensional Ising model on the simple cubic lattice, a typical example whose exact solution remains unknown. Obviously, the explicit expression of the density function of Fisher zeros has not yet been obtained. As pointed out by Domb (see Sec. \uppercase\expandafter{\romannumeral3} of Ref. \cite{RN451}), the series of terms consistent in sign leads to a dominant singularity on the positive real axis (perhaps the physical critical point), while in the case that coefficients alternate in sign the dominant singularity lies on the negative real axis. This is consistent with our findings of this paper. Domb also concluded that, the alternating signs in the low-temperature series of three-dimensional Ising model lead to spurious non-physical singularities, which mask the true critical behaviour (see Sec. \uppercase\expandafter{\romannumeral1} of Ref. \cite{RN451}). 
Here we briefly reexamine this statement.
Considering the zero field case, we still use $z=e^{2\beta J}$ as the variable in the low-temperature series $\sum\limits_n a_n z^n$, and use the subgraph expansion on the simple cubic lattice to verify the coefficient $a_n$. The first terms can be obtained
\begin{align}
&~~~\left\{ \mu_6, \mu_{10}, \mu_{12}, \mu_{14}, \mu_{16}, \cdots \right\}  \nonumber \\
& = \left\{ N, 3N, \frac{N\left( N - 7 \right)}{2}, 15N, 3{N^2} - 33N, \cdots \right\}   \label{eq93}
\end{align}
and
\begin{equation}
\left\{ a_6, a_{10}, a_{12}, a_{14}, a_{16}, \cdots \right\} = \left\{ 1, 3, -\frac{7}{2}, 15, -33, \cdots \right\}.     \label{eq94}
\end{equation}
Note that the second column of Table \uppercase\expandafter{\romannumeral1} of Ref. \cite{RN425} listed the sequence $\left\{2na_n\right\}$, and one can also obtain $\left\{a_n\right\}$ from the values therein. It seems that, the accumulation point with the smallest modulus (the dominant singularity) is on the negative real axis, but there is still a physical critical point on the interval $\left[ 0,1 \right]$. Therefore, the distribution of Fisher zeros of the simple cubic lattice Ising model is a very interesting but challenging problem. Further exploration of this issue is warranted.

\blue{Finally, we note that the Fisher zeros can also be applied to quantum phase transition. In the case of transverse field Ising model, the Fisher zeros can also be used to determine the quantum critical point. For example, Ref. \cite{RN503} studied the Fisher zeros of the analytically continued one-dimensional transverse field Ising model. It demonstrated that the lines of Fisher zeros evolve smoothly as the transverse field is tuned, and a qualitative change identifies the quantum critical point. This finding suggested that dependence of critical temperature on the transverse field can be determined from the accumulation points of Fisher zeros. Therefore, applying of Fisher zeros to the low-temperature series of quantum spin systems deserves further study.}

\begin{acknowledgments}
This work was supported by Guangdong Provincial Quantum Science Strategic Initiative (Grants No. GDZX2203001 and No. GDZX2403001), National Natural Science Foundation of China (Grants No. 12474489 and No. 12474228), the Research Funding for Outbound Postdoctoral Fellows in Shenzhen (Grant No. SZRCXM2401006), Shenzhen Fundamental Research Program (Grant No. JCYJ20240813153139050), and the Innovation Program for Quantum Science and Technology (Grant No. 2021ZD0302300).
\end{acknowledgments}

\end{document}